\documentclass[epj]{svjour}
\usepackage{graphics,amssymb,graphicx,bm,color}
\newcommand\ETO{$\rm Er_2Ti_2O_7$}
\newcommand\GTO{$\rm Gd_2Ti_2O_7$}

\newcommand\pyro{pyrochlore}
\newcommand\PRB[3]{Phys. Rev. B {\bf {#1}}, ({#3}) {#2}}	
\newcommand\PCM[3]{J. Phys.: Cond. Matter {\bf {#1}}, ({#3}) {#2}}

\newcommand\PPRL[3]{Phys. Rev. Lett. {\bf {#1}}, ({#3}) {#2}}				 
\newcommand\PhysB[3]{Physica B {\bf {#1}}, ({#3}) {#2}}
\newcommand\JPCS[3]{J. Phys.: Conf. Series {\bf {#1}}, ({#3}) {#2}}
\newcommand\MH{Magnetisation}
\newcommand\mH{magnetisation}
\begin{document}
\title{Low-temperature magnetisation process in the cubic pyrochlore quantum antiferromagnet, {$\rm Er_2Ti_2O_7$}}
\author{O. A. Petrenko, M. R. Lees and G. Balakrishnan}
\institute{University of Warwick, Department of Physics, Coventry, CV4~7AL, UK}
\date{Received: date / Revised version: date}
\abstract{We report low-temperature susceptibility and \mH\ data for the cubic \pyro\ \ETO.
By performing the measurements on single crystals we are able to establish the degree of magnetic anisotropy present in this compound and to determine the critical values for magnetic field induced transitions below the magnetic ordering temperature of 1.2~K.
We also present a magnetic $H-T$ phase diagram of this quantum XY antiferromagnet for different directions of an applied field.
}
\maketitle
\section{Introduction}
\label{intro}
The cubic \pyro\ oxides, $\rm A_2B_2O_7$, play a central role in the research of frustrated magnetism~\cite{Pyros}.
One of the members of the titanium rare-earth \pyro s, \ETO, represents a rare example of a magnetic system where the effect of a selection of a particular ground state out of a degenerate manifold through quantum and thermal fluctuations (the so called order-by-disorder, OBD, effect) can be studied in detail.  
Although suspected of being important in \ETO\ and first proposed for consideration about a decade ago~\cite{Champion_PRB_2003,Champion_2004}, the OBD mechanism has only very recently been proven to be solely responsible for the establishment of a non coplanar ground state at low temperature in this compound.
This has been achieved by the development of the appropriate theory~\cite{Zhitomirsky_PRL_2012,Savary_PRL_2012} and via comparison with an extensive set of experimental data, which includes neutron scattering~\cite{Champion_PRB_2003,Savary_PRL_2012,Pool,Ruff_PRL_2008}, magnetisation~\cite{Bramwell_JPCM_2000}, ca\-lo\-rimetry~\cite{Ruff_PRL_2008,Blote_Physica_1969,Siddrathan_PRL_1999,Sosin_PRB_2010} and ESR~\cite{Sosin_PRB_2010} data. 
Surprisingly, a comprehensive set of low-temperature magnetisation data, $M(H,T)$, are not available in the literature (apart from our early report~\cite{Petrenko_JPCM_2011} for $H \parallel [111]$).
In this paper we report the results of the magnetisation measurements for two other principal directions of the applied magnetic field for cubic symmetry, $H \parallel [100]$ and $H \parallel [110]$. 

\section{Experimental procedures}
Single crystal samples of the \ETO\ were grown by the floating zone technique using an infrared image furnace~\cite{Balakrishnan_JPCM_1998}.
The high quality of the samples was confirmed by previous calorimetry measurements~\cite{Sosin_PRB_2010}, which showed a sharp peak in the heat capacity at the magnetic ordering temperature.
We confirmed the reproducibility of the obtained data by repeating the measurements on a second sample, prepared and aligned separately.
The principal axes of the samples were determined using the X-ray diffraction Laue technique; the crystals were aligned to within an accuracy of 2$^{\circ}$.
The plate-like shape of the samples used (with the magnetic field applied in the sample plane) has insured minimal de\mH\ effects - the estimated difference between $H_{\rm applied}$ and $H_{\rm internal}$ did not exceed 6-7\%. 
\MH\ measurements were made down to 0.5~K in applied magnetic fields of up to 70~kOe using a Quantum Design MPMS SQuID magnetometer along with an {\it i}Quantum $^3$He insert~\cite{Shirakawa_2004_JMMM}.
The \mH\ was measured both as a function of temperature in a constant magnetic field and as a function of applied field at constant temperature. 

\section{Experimental results and discussion}
\begin{figure*}
\begin{center}
\includegraphics[width=0.8\columnwidth]{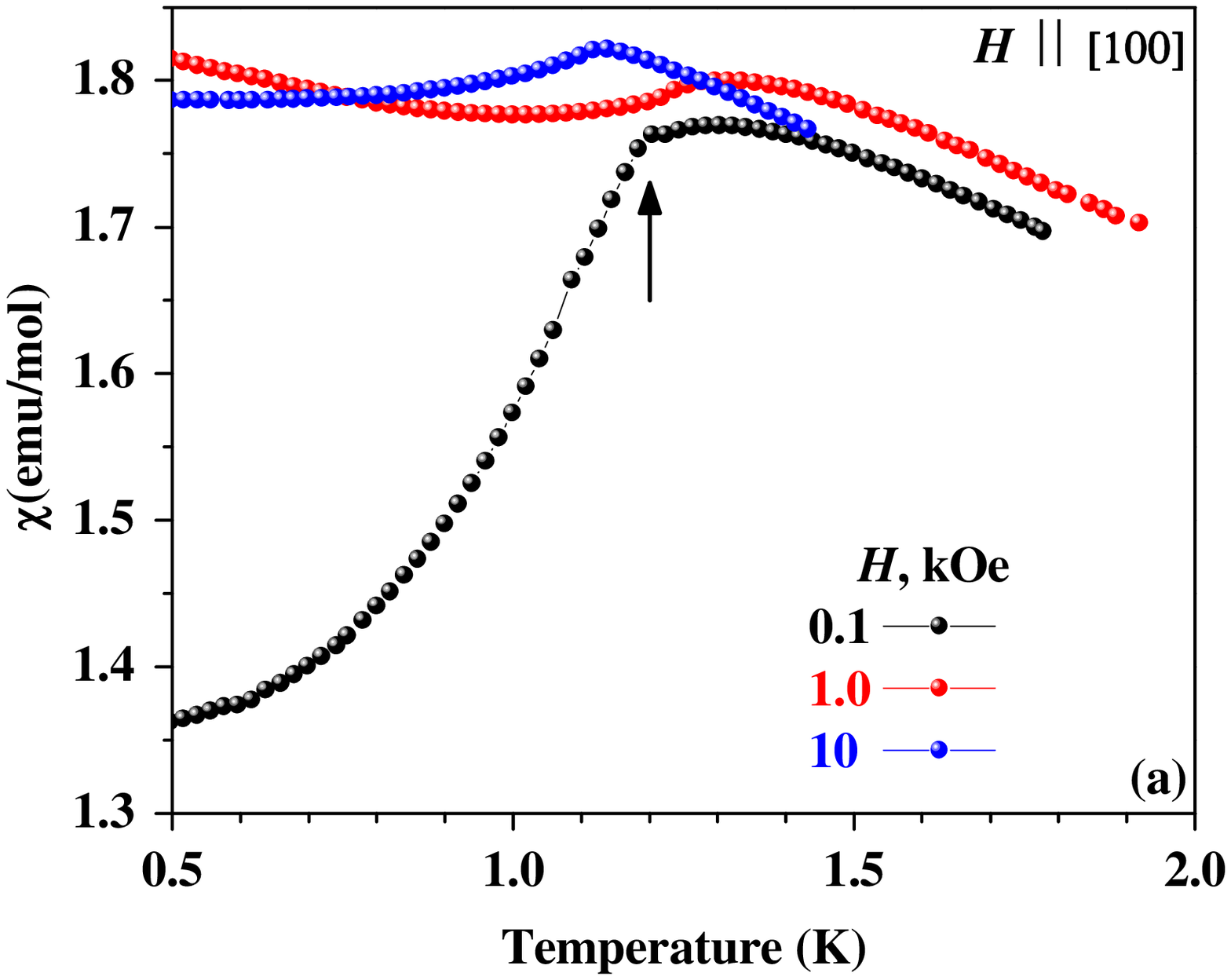}
\includegraphics[width=0.8\columnwidth]{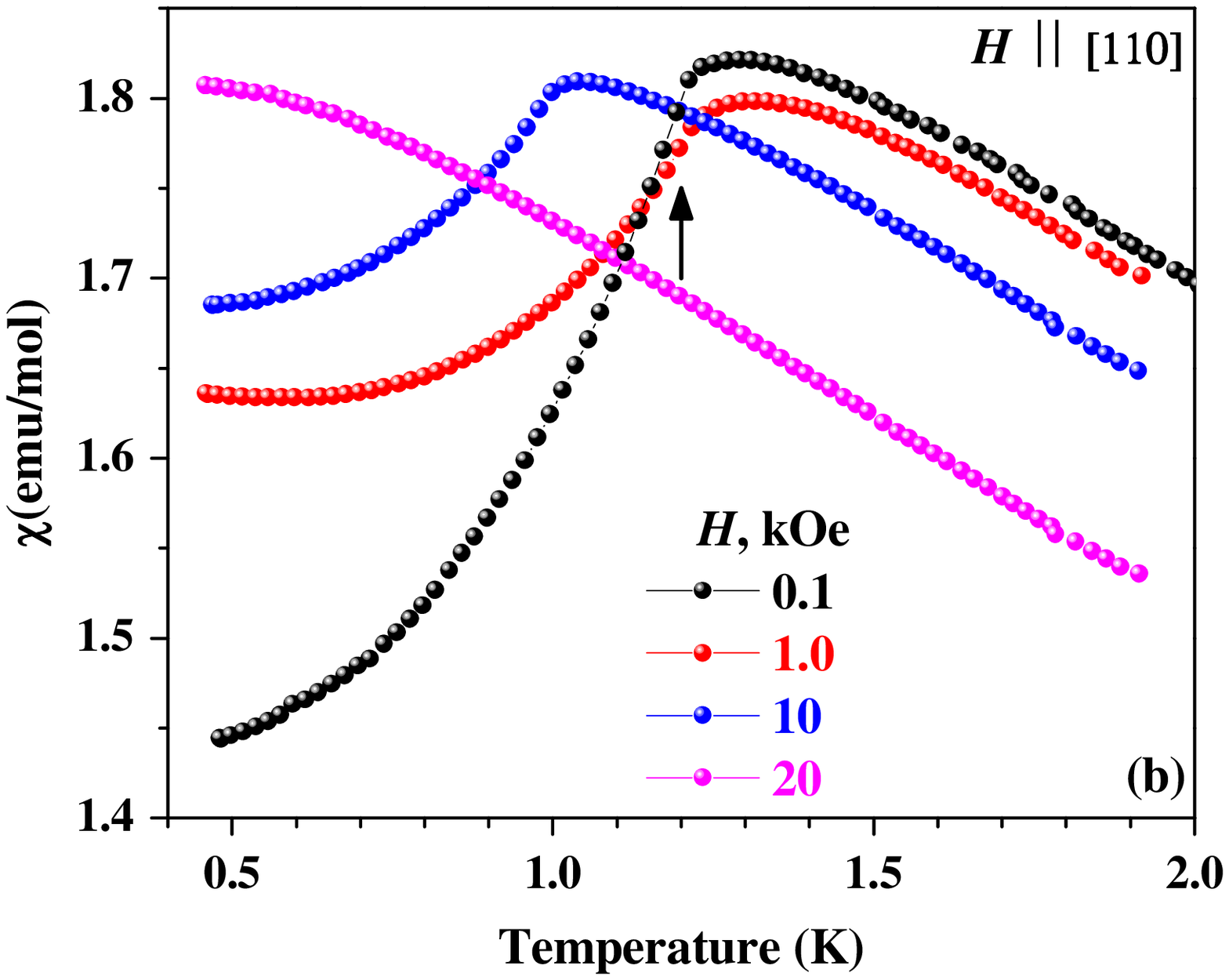}
\caption{Temperature dependence of the magnetic susceptibility of \ETO\ in different applied fields for $H \parallel [100]$ (a) and $H \parallel [110]$ (b).
		The arrows indicate the anomaly associated with the ordering temperature $T_N=1.20$~K.}
\label{Fig1_MT} 
\end{center}
\end{figure*}
\begin{figure}
\begin{center}
\includegraphics[width=0.8\columnwidth]{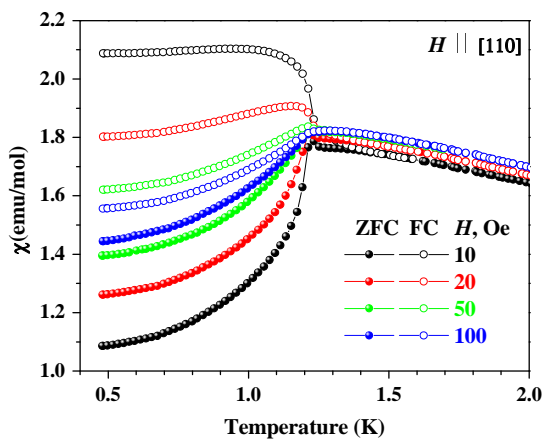}
\caption{Temperature dependence of the magnetic susceptibility of \ETO\ in weak fields applied along the $[110]$ direction.
The results for zero field cooled (ZFC) and field cooled (FC) measurements are shown in solid and open symbols respectively.}
\label{Fig2_MT_low}
\end{center}
\end{figure}
\begin{figure*}
\begin{center}
\includegraphics[width=0.8\columnwidth]{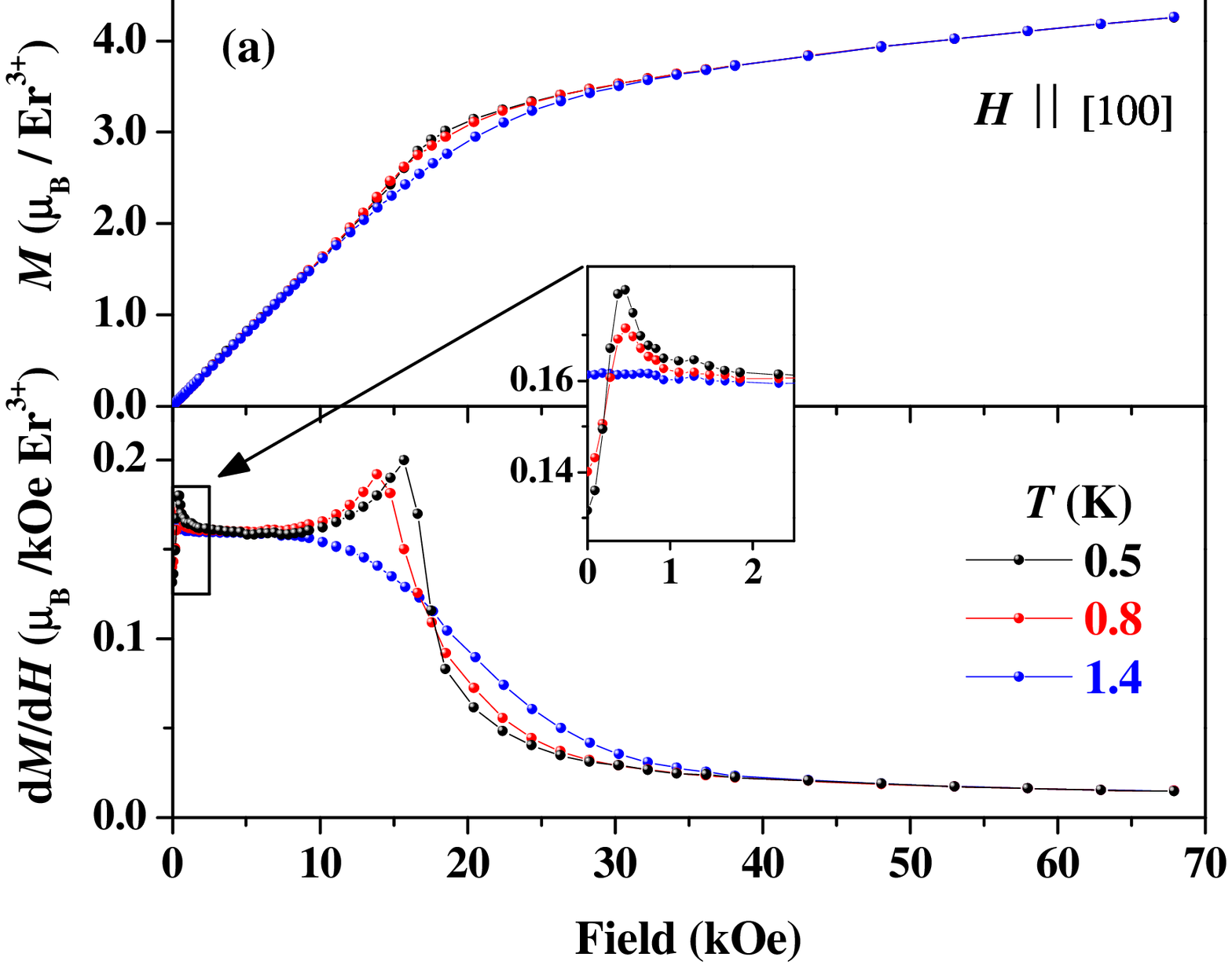}
\includegraphics[width=0.8\columnwidth]{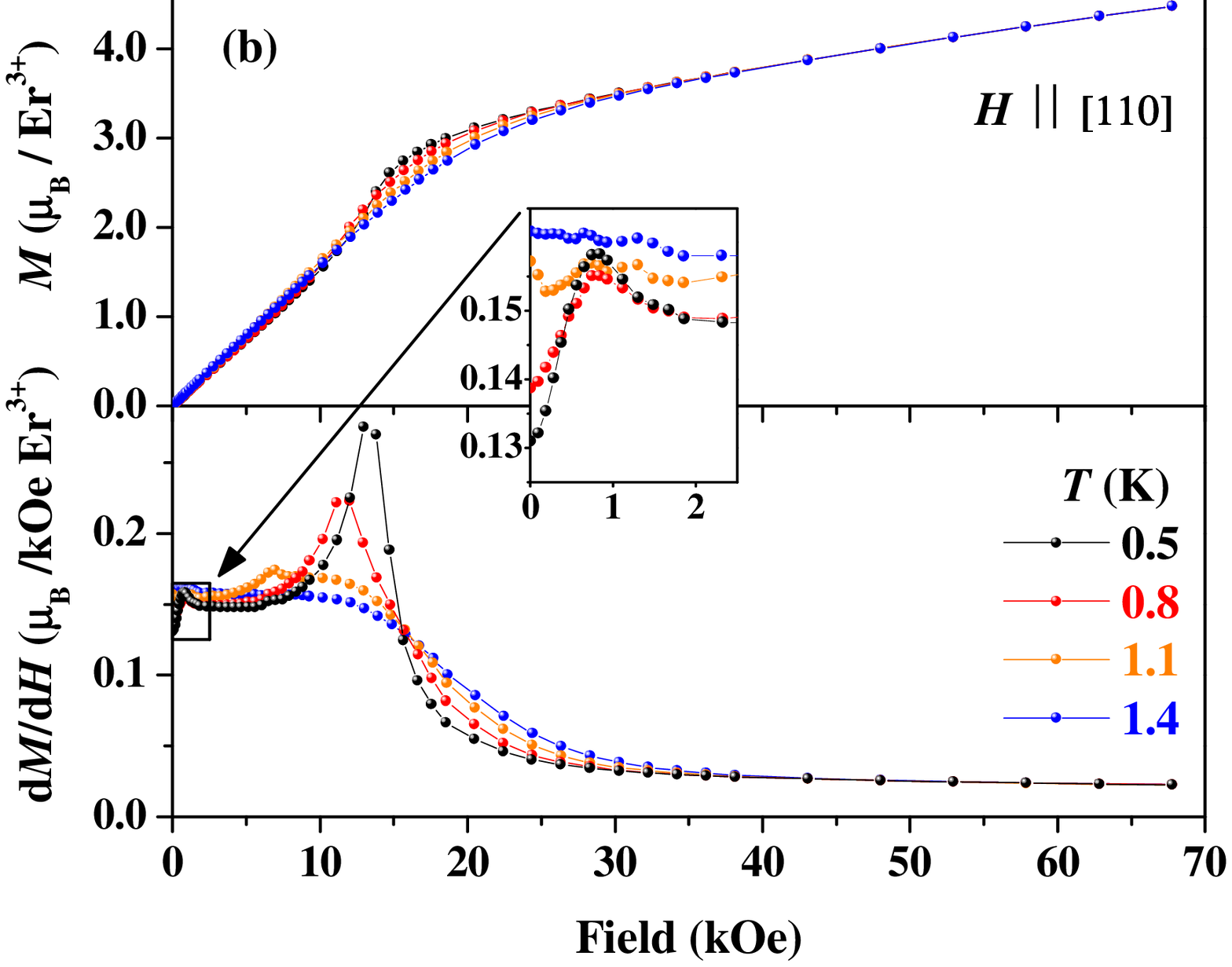}
\caption{\MH\ (upper panels) and $dM/dH$ (lower panels) curves at different temperatures for $H \parallel [100]$ (a) and  $H \parallel [110]$ (b).
		The insets emphasise the non-monotonic behaviour of the magnetisation derivatives $dM/dH$ in low applied field for $T<T_N$.}
\label{Fig3_MH}
\end{center}
\end{figure*}

The results of the temperature and field dependence of \mH\ for $H \parallel [100]$ and $H \parallel [110]$ are presented in Figures~\ref{Fig1_MT}, \ref{Fig2_MT_low} and \ref{Fig3_MH}.

For both field directions, the susceptibility measured in a low field of 0.1~kOe decreases rapidly below the ordering temperature, $T_N=1.20$~K.
$T_N$ is marked by a sharp anomaly in $\chi(T)$, similar to that reported previously~\cite{Petrenko_JPCM_2011} for $H \parallel [111]$.
In lower fields, an anomaly associated with $T_N$ is clearly visible just below the maximum in $\chi(T)$.
In higher fields, $T_N$ is well defined by the maximum of the temperature derivative of the product $\chi(T) T$.
The value of $T_N$ determined in this manner is in agreement with previous calorimetry data~\cite{Ruff_PRL_2008,Blote_Physica_1969,Siddrathan_PRL_1999,Sosin_PRB_2010}, 
as well as susceptibility data obtained on a powder sample~\cite{Blote_Physica_1969}.
For the measurement in a larger applied field of 1.0~kOe, the sharp anomaly is replaced by a much smoother variation in the $\chi(T)$ curves and the decrease at $T < 1.2$~K is significantly less pronounced for $H \parallel [110]$, while for $H \parallel [100]$ the susceptibility actually increases with decreasing temperature below 1~K.
A further tenfold increase in the applied field causes the maximum in $\chi(T)$ to shift to 1.04~K for $H \parallel [110]$, while for $H \parallel [100]$ the corresponding shift is rather small (to 1.14~K).
For the measurements performed in a field of 20~kOe and above, the $\chi(T)$ curves vary smoothly with temperature showing no sign of a magnetic transition over the entire temperature range covered (curve shown in Fig.~\ref{Fig1_MT}b for $H \parallel [110]$).

For the higher applied fields, no appreciable history dependence is observed in the measured $\chi(T)$ curves.
For the lower fields, the susceptibility measurements show a nearly two-fold difference in the values of $\chi(T=0.5$~K) depending on the measurement protocol, zero field cooled warming or field cooled warming.
The corresponding data for the $[110]$ direction are displayed in Fig.~\ref{Fig2_MT_low}.
For $H \parallel [100]$, the data for low-field susceptibility look very similar to those shown in Fig.~\ref{Fig2_MT_low} and to the low-field data for the $[111]$ direction~\cite{Petrenko_JPCM_2011}.
For all of them the biggest difference between FC and ZFC regimes is observed in the lowest fields.
Note, that the accuracy of the absolute value of $\chi (T)$ measured in very low fields is rather limited due to an ever present trapped field in the superconducting solenoid.

Fig.~\ref{Fig3_MH} shows that for both directions of the applied field studied, the \mH\ increases smoothly with field for $T>T_N$.
The corresponding $dM/dH$ curves demonstrate a very gradual transformation from a rather large gradient of $\sim 0.15$~$\mu_B$/kOe seen in lower fields to a much smaller, but still considerable, gradient of $\sim$0.015 to 0.025 $\mu_B$/kOe in higher fields.
Our preliminary data for high-field magnetisation~\cite{HighField} taken at 1.5~K suggest that the magnetisation continues to increase in fields as high as 600~kOe and reaches 9 to 10~$\mu_B$ per Er$^{3+}$ ion in the highest field.

At lower temperatures there is a much more pronounced change in the slope of $M(H)$ at and immediately above a critical field $H_c$.
Most of the temperature dependence in the \mH\ curves manifests itself in how fast the change from one gradient to another occurs around this critical field $H_c$, while the high field part of the \mH\ remains practically temperature independent. 
Below $T_N$, the \mH\ curves develop a characteristic $S$ shape with the corresponding $dM/dH$ curves demonstrating a clear maximum, which provides a natural way of defining the critical field.
Defined in this way, $H_c$ at $T=0.5$~K amounts to 15.7 and 13.5~kOe for $H \parallel [100]$ and  $H \parallel [110]$ respectively.
The maximum of the $dM/dH$ curve is much higher for the $[110]$ direction, therefore the associated $S$ shape is much more pronounced.
No significant hysteresis has been observed at any temperature in any of the \mH\ curves for increasing and decreasing fields.

If one assumes a simple XY model in which the four identical magnetic moments on each tetrahedron are always confined to a local $(111)$ plane, then for $H \parallel [100]$ the four ions contribute equally to the \mH\ and the maximum field-induced moment per Er$^{3+}$ ion is restricted to $\mu_{100}^{\rm max}=\mu \sqrt{\frac{2}{3}}$, where $\mu$ is the value of the Er$^{3+}$ magnetic moment in \ETO.
Taking $\mu=3.25(9)$~$\mu_B$ from a spherical neutron polarimetry study~\cite{Pool}, the calculated value for $\mu_{100}^{\rm max}$ is 2.65(7)~$\mu_B$, while the experimentally observed value of $M_{100}(H=H_c)$ is 2.62(2)~$\mu_B$.
For $H \parallel [110]$, two out of the four ions can contribute their full moment to the \mH\, as the field is applied within their easy-plane, while the contribution from the other two ions is restricted to $\mu/\sqrt{3}$ each.
Therefore the maximum field-induced moment per Er$^{3+}$ ion is calculated as $\mu_{110}^{\rm max}=\mu \frac{1+\sqrt{3}}{2\sqrt{3}}=2.56(7)$$\mu_B$, while the experimentally observed value of $M_{110}(H=H_c)$ is 2.27(2)~$\mu_B$.
The agreement between the calculated and observed magnetisation values at a critical field is perfect for $H \parallel [100]$, but less satisfactory for $H \parallel [110]$, which is not entirely surprising given the simplified nature of the model assumed.
\begin{figure}
\begin{center}
\includegraphics[width=0.9\columnwidth]{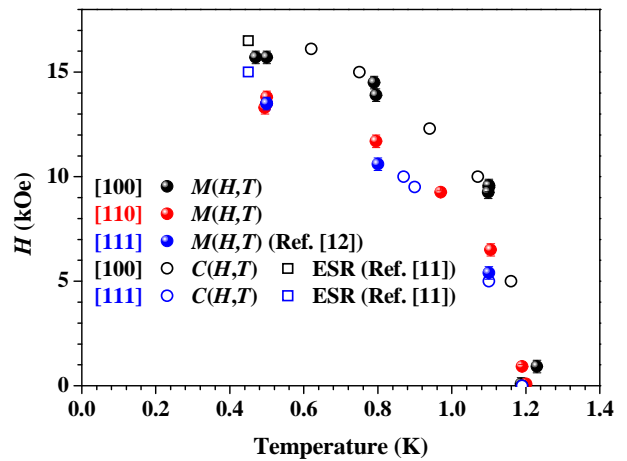}
\caption{Magnetic phase diagram of \ETO\ for different directions of an applied field.
		Solid symbols are from \mH\ measurements, while the open symbols correspond to the heat capacity ($\bigcirc$) and ESR ($\square$) measurements~\cite{Sosin_PRB_2010}.}
\label{Fig4_PhD}
\end{center}
\end{figure}

The insets in Fig.~\ref{Fig3_MH} emphasise the field dependence of the \mH\ derivative $dM/dH$ in lower fields.
Below $T_N$, nonlinear field dependence is observed for both $H \parallel [100]$ and  $H \parallel [110]$ with a clear maximum appearing at around 0.5-0.8~kOe.
This behaviour is likely to be associated with the movement of magnetic domain walls in the sample in the magnetically ordered state.
Remarkably, no significant non-linearity in the $dM/dH$ has been previously observed~\cite{Petrenko_JPCM_2011} for $H \parallel [111]$.

Fig.~\ref{Fig4_PhD} shows the resulting magnetic phase diagram of \ETO, in which the magnetisation data from this paper ($H \parallel [100]$ and $H \parallel [110]$) have been combined with the data from our previous paper~\cite{Petrenko_JPCM_2011} for $H \parallel [111]$ as well as the heat capacity and ESR data~\cite{Sosin_PRB_2010}.
A significant anisotropy of the magnetic behaviour of \ETO\ becomes obvious after considering this Figure, with the [100] direction demonstrating a shift to higher fields compared to the rather similar results for the other two directions.
Overall, the agreement between the magnetisation and heat capacity data is reasonably good, but it would be considerably poorer if the temperatures of the maxima of the $\chi(T)$ curves were taken to correspond to the transition temperature and were included in the plot.
For example, in a field of 10~kOe for $H \parallel [100]$, the transition in the $C(T)$ is observed at 1.07~K, while the $\chi(T)$ has a maximum at 1.13~K, suggesting clearly that the maximum occurs in the paramagnetic phase.
Rather than corresponding to a maximum in $\chi(T)$, the transition is more likely to be marked by a maximum slope in the $\chi(T)$ curves.

The magnetic phase diagram of the $XY$ antiferromagnet \ETO\ looks significantly less complex than what has been observed in the Heisenberg \pyro\ antiferromagnet \GTO, where multiple field-induced transition have been seen~\cite{GTO}, resulting in the presence of as many as five different magnetically ordered phases for some directions of an applied field.
A common feature for these two \pyro\ systems is a rearrangement of the magnetic domains by relatively modest fields.  

\section{Summary}
Significant progress in the understanding of the properties of \ETO\ has been achieved recently through the theoretical considerations~\cite{Zhitomirsky_PRL_2012,Savary_PRL_2012,McClarty_JPCS_2009}, although not every issue has been resolved to date, as there are alternative descriptions~\cite{Briffa_PRB_2011} of the magnetism in this compound.
One paper~\cite{Zhitomirsky_PRL_2012} explicitly called for additional measurements of the angular dependence of the critical field $H_c$ at low temperatures to allow for a more precise identification of the microscopic parameters of the model used in the case of \ETO.
The authors hope that the data presented in this paper will serve this purpose.

Note added. Recently, low-temperature magnetisation measurements~\cite{Bonville_JPCM_2013} have been reported for \ETO\ for the three high symmetry directions of applied field.
The results of these measurements performed at 0.13 to 0.175~K are largely in agreement with our higher-temperature measurements, although the values of critical fields are slightly higher at 18 and 16~kOe for $H \parallel [100]$ and $H \parallel [110]$ respectively.

\section{Acknowledgements}
The magnetometer used in this research was obtained through the Science City Advanced Materials project: Creating and Characterising Next Generation Advanced Materials project, with support from Advantage West Midlands (AWM) and part funded by the European Regional Development Fund (ERDF).
The authors also acknowledge financial support from the EPSRC, United Kingdom.

\end{document}